# Behavior of the local mode's potential in BaTiO$_3$ studied by Effective Hamiltonian numerical simulations

MANUEL I. MARQUÉS [a], AND JULIO A. GONZALO [a]

[a] *Departamento de Física de Materiales C-IV, Facultad de Ciencias,*

*Universidad Autónoma de Madrid, 28049 Madrid, Spain*

The behavior of the local mode's potential energy in the unit cell of BaTiO$_3$ is studied by effective Hamiltonian numerical simulations. The study is focused on the evolution of the potential's minimum felt by the Ti ion. At the paraelectric phase, the single instantaneous minimum is located at one of eight off-centered regions at the body diagonals of the unit cell. The study shows graphically how the soft mode behavior of the perovskite comes from the average response of the Ti ion to the potential changes in the unit cell, showing that the single minimum of the time-averaged potential is centered at the paraelectric phase and off-centered at the ferroelectric phase.

RUNNING HEAD: BaTiO$_3$ local mode's potential

## INTRODUCTION

The nature of ferroelectric phase transitions in perovskites has been intensely investigated over many decades. The first microscopic theory for BaTiO$_3$, developed by Mason and Matthias, dates back to 1948 [1]. It did take into account the central role played by the Ti ion, seen as jumping between a set of off-centered potential minima in the unit cell. Later, Comes et al. [2], proposed a model where the Ti ion is located at one of the eight positions shifted from the center along the body diagonals of the unit cell. This model, called the 8-site model, was based on a pure order-disorder mechanism. At low temperature the Ti ion moves preferentially between the four sites in the y-z plane leading to an effective displacement on a given x direction. However, pure order-disorder models were unable to explain the large value of the Curie-Weiss constant and the small value of the transition entropy [3]. A totally different new theory came in 1960 from Cochran [4]. It is a so-called displacive theory: The potential for the Ti motion in the paraelectric phase must exhibit a minimum at the center of the unit cell, and it must change at the critical temperature, turning to be off-centered and resulting in a spontaneous polarization different from zero. Note that in this theory the potential energy is temperature dependent. The success of this theory came from an, in principle, observable experimental implication: the existence of soft modes. When the potential structure at the unit cell becomes unstable and changes at the phase transition some normal modes connected to a given restoring force must soften so as to allow the phase transition. Inelastic neutron scattering [5] and Raman spectroscopy measurements [6] revealed the softening of the transverse optic phonon near the Brillouin zone center. This experimental confirmation was considered as a definitive back up for the displacive theory.

However, the order-disorder mechanism was never completely discarded, mainly due to experimental results obtained by means of other experimental techniques, such as diffuse x-ray scattering. Experimental results as the ones obtained by Comes et al. [2] show the presence of high temperature ($T>T_c$) correlated clusters of quasistatic off-center Ti ions, as expected for an order-disorder transition. In fact, during the last decade, structural experiments based on extended x-ray absorption fine structure, nuclear magnetic resonance and correlation pair distribution function have given results for the local structure of oxide ferroelectrics which are closer to the order-disorder expectatives than to the displacive theory presumptions. These results deviate from the ones previously obtained with neutron scattering or Ramman spectroscopies. As an example, the following results were obtained in a recent x-ray absorption fine structure (XAFS) experiment on $BaTiO_3$ [7]: (i) Local displacements of the Ti atoms from their cubic sites in the tetragonal phase were found to be approximately rhombohedral, deviating from (111) directions by 12º toward the c-axis. (ii) At the cubic phase local displacements are more closely directed along the (111) direction than along the (100) direction. (iii) The magnitude of the Ti atom displacement is always different from zero and it remains almost constant with a slight change of 13% from 35K to 590K (iv) The eight (111) equivalent displacements remain for temperatures even 350 ºC above $T_c$.

Half-way between the neutron scattering experiments and the XAFS studies we find the new results obtained by NMR [8]. These experiments are claimed to be evidence for an old idea that has been around in the ferroelectrics literature for some time: the coexistence of order-disorder and displacive components in Barium Titanate [9]. This experiment indicates that, up to 35ºC above $T_c$, the Ti atoms remain displaced from their cubic sites but with a local tetragonal type distortion. The paraelectric behavior well above $T_c$ is due to the absence of long-range correlation among local tetragonal displacements. By cooling the system a displacive component is found when passing through $T_c$, as the electric field gradient tensor increases by a factor of 100.

The question is clear: *Why all these experiments do not coincide?* Simply, neutron scattering experiments seem to point toward the displacive theory, XAFS measurements appear to point toward the 8-site model, with an orientation shift of the displacements at the tetragonal phase [10], and NMR seems to support a coexistence between both, displacive and order-disorder components.

A possible answer to this question was given some time ago: the results depend on the sampling time scale of the experiment. Basically, it was shown that the observability of a double-well potential is dependent on the time scale of the probe [11]. In fact, the various collective particle dynamics are governed by different time and length scales and the coexistence of these various time scales may result in the coexistence of order-disorder and displacive observed behavior [12, 13]. This idea has been used as a hypothesis on a recent paper [10] to explain the difference between the NMR and XAFS results. NMR averages dynamic displacements that vary in times of the order of $10^{-8}$s, while XAFS has an averaging time of less than $10^{-15}$s. XAFS measures the instantaneous local structure and detects the 8-site (111) displacements. On the other hand NMR, while being fast enough to measure a displacement different from zero it is not capable of measuring the (111) displacements and performs an average resulting in a (100) tetragonal displacement. The displacive component measured by NMR is actually a shift on the displacements of the Ti ions and not a change on their magnitude.

In this paper we will study straightforwardly and in a transparent way whether this hypothesis of different time scale for the displacive and order-disorder components is true or not. By "ab-initio" effective Hamiltonian numerical simulations we will check for the first time if the instantaneous measurements detect 8-site (111) displacements and if time-averaged methods, such as Raman or neutron scattering, detect the net displacive component behavior coming from a change from a centered to a non-centered potential.

**MODEL AND COMPUTATIONAL DETAILS**

To get a clear idea about the nature of the phase transition in $BaTiO_3$ we study the behavior of the potential energy in the cubic cell of $BaTiO_3$ at the paraelectric and ferroelectric phases. In particular, it is possible to study the value of the potential energy $V(u_i,t)$ at a particular local cell i and at a given time t as the local mode displacement $u_i=(ux_i,uy_i,uz_i)$ (or local polarization $p_i$) takes different values in the cell.

The potential energy is going to be calculated using first-principles numerical simulations. It is important to stress that the calculated local mode potential energy takes into account the effect of all unit cells in the sample. The model used is the same effective Hamiltonian specified by ultrasoft pseudopotentials introduced by Zhong et al. [9]. This effective Hamiltonian takes into account a local soft-mode self-energy containing intersite interactions to quartic anharmonic order, a long range dipole-dipole coupling, a short range correction to the intersite coupling going up to third neighbors, a harmonic elastic energy and finally, an anharmonic strain-soft mode coupling. Details about the model and values of the parameters may be found in [14]. The behavior of the Hamiltonian at different temperatures is studied using standard Monte Carlo techniques. This model successfully reproduces the phase transition sequence for $BaTiO_3$ although with critical temperatures differing from the ones observed experimentally. In particular we are going to study the system at T=400K (corresponding to the cubic-paraelectric phase) and at T=250K (corresponding to the tetragonal-ferroelectric phase). The study is focused on $BaTiO_3$ cubic systems with 9×9×9 unit cells and with periodic boundary conditions at a constant pressure P=-4.8 GPa.

We are interested in performing a comparison of the value for the potential energy inside the unit cell (in particular the position of the minima) when using an instantaneous experimental technique (such as XAFS) and when using a time-average technique (such as neutron scattering). To do so, we are going to perform two kinds of simulations:

(i) Once the system is thermalized at a given temperature we consider each instantaneous equilibrium configuration of the system focusing our attention on a given cell i. At that particular equilibrium configuration t we measure $V(u_i,t)$ (which is given by the sum of all terms in the Hamiltonian) for different values of $u_i$ all over the unit cell. Due to the symmetry of the potential self-energy we are going to find a single minimum $V(u^m_{i,t},t)$ corresponding to a particular displacement inside the unit cell that we will denote as $u^m_{i,t}$.

If the hypothesis we are checking is true, after collecting $u^m_{i,t}$ for different $N_{conf}$ instantaneous configurations, we will find minima located at the 8 positions shifted from the center along the body diagonals of the unit cell when T=400K and just at 4 of the 8 positions (the ones in the y-z plane) when measuring at the ferroelectric phase, T=250K.

(ii) Next, we will consider a time-average measuring technique. To do so, we are going to calculate not the instantaneous potential energy, but the time-averaged one

$$\overline{V}(u_i) = \frac{1}{N_{conf}} \sum_{t=1}^{N_{conf}} V(u_i,t) \tag{1}$$

Again, if the hypothesis of a displacive component coming from the long time scales holds true we should find an averaged potential energy with a single minimum at the center of the unit cell (i.e. at $u_i=0$) when T=400K and with a off-centered minimum displaced at the x-axis when T=250K. In all cases, we are going to consider $N_{conf}$=20000.

**INSTANTANEOUS MEASUREMENTS RESULTS**

The position of the instantaneous minima collected after considering 20000 different equilibrium configurations are plotted on Fig.1a for T=400K and Fig.1b for T=250K. In order to analyze the results with more detail we have plotted not all minima $\{u^m_{i,t}\}$ but only the ones corresponding to the deeper values of the potential energy, i.e. those fulfilling

$$V(u^m_{i,t},t) < (V^m + 0.5|V^m|) \tag{2}$$

being $V^m$ the deepest value of the potential energy found after considering the 20000 configurations.

Fig.1a shows how instantaneous measurements (i.e. measurements made in $10^{-15}$s) detect the correctness of the early 8-site view by Comes et al. [2]. Potential minima for the Ti ion are located at eight positions shifted from the center along the body diagonals of the unit cell at the paraelectric phase. Note how the eight off-centered regions are clearly shown in the three projections over the lateral planes. On the other hand, Fig1b indicates that in the ferroelectric phase deeper minima are located in an extended surface formed mostly by only four of the eight previously reported regions. The surface is formed toward the direction given by the axis of the local dipole. This results is in agreement with XAFS measurements [7], where, as already explained, local displacements of the Ti atoms from their cubic sites in the tetragonal phase are found to be rombohedral, but shifted from the (111) directions by approximately 12º towards the c-axis. So, the first part of the hypothesis has been checked: ultra-fast, instantaneous measurements, find a mechanism very similar to the seminal order-disorder 8-site model proposed by Comes et al. [2]. Now we should check if much slower time scale measurements, such as neutron crystallography, detect the displacive mechanism.

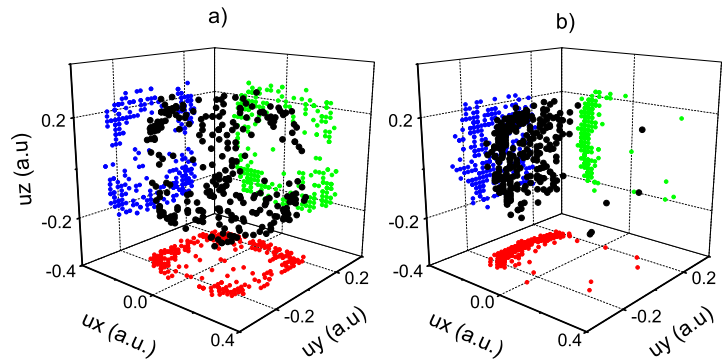

FIGURE 1. Position (in atomic units) of all the instantaneous local minima fulfilling Eq.2 after considering 20000 different configurations. Temperature is equal to (a) T=400K corresponding to the paraelectric phase and (b) T=250K corresponding to the ferroelectric phase. Projections over the plane are represented with green (light gray), red (gray), and blue (dark gray) points.

**TIME-AVERAGED MEASUREMENTS RESULTS**

Results for the paraelectric and the ferroelectric phases are presented in Fig. 2a and 2b respectively. We represent the region of the local cell where the averaged potential is deeper, i.e., we plot $u^m_i$, that is the position of the total minimum of the averaged potential, together with all local modes $u_i$ fulfilling:

$$\overline{V}(u_i) < \left[\overline{V}(u^m_i) + f\,|\overline{V}(u^m_i)|\right] \tag{3}$$

being f=50 for the paraelectric phase and f=0.5 for the ferroelectric phase.

We find that $u^m_i \sim 0$ in the paraelectric phase and $u^m_i \neq 0$ at the ferroelectric phase.

So, the question about why $BaTiO_3$ behaves as if the Ti atom is centered at the paraelectric phase and displaced at the ferroelectric phase is now answered considering a measurement of the long term averaged behavior of the potential. It means that the second part of the hypothesis is also true: If we make an experiment of a ferroelectric transition using slow time scale measurements we are going to find a displacive behavior as the one proposed by Cochran [4].

In order to get a more clear idea about the behavior of the average potential energy inside the cell, it is useful to plot the averaged potential versus $u_i$ for $u_i=(ux_i,0,0)$.

Results are presented in Fig.3 for both, the paraelectric and ferroelectric phases. Note how at the paraelectric phase the minimum is almost zero and is located at the center of the unit cell and how, in the ferroelectric phase, the minimum is negative and it is displaced from the center of the unit cell.

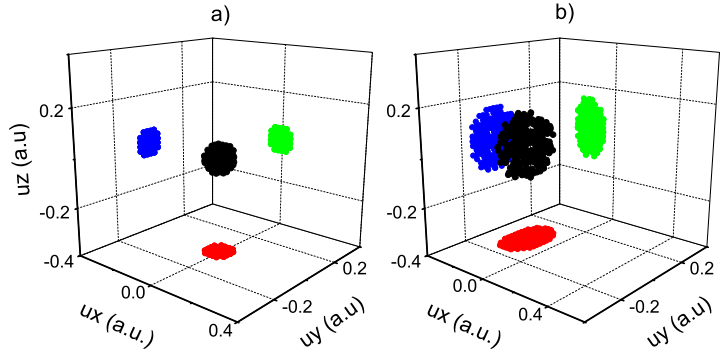

FIGURE 2. Position (in atomic units) of all local modes fulfilling Eq. 3 after considering 20000 configurations for (a) T=400 K corresponding to the paraelectric phase and (b) T=250K corresponding to the ferroelectric phase. Projections over the planes are represented with green (light gray), red (gray), and blue (dark gray) points.

**CONCLUSIONS**

In this paper we have shown, using a new and fairly intuitive method, that the hypothesis of a different time scale for the displacive and order-disorder components on Barium Titanate is true. Using "ab-initio" effective Hamiltonian numerical simulations we have checked, for first time, how instantaneous measurements detect 8-site (111) displacements and time-averaged methods detect the effective displacive behavior coming up from the change from a centered to a non-centered potential.

Our calculations show clearly how the long standing problem of the displacive vs. the order-disorder character of the ferroelectric transition on $BaTiO_3$ is just a measurement problem. The instantaneous mechanism of the phase transition in perovskites corresponds to a mechanism very similar to the 8-site model, but results obtained when measuring with slow time scale devices are rather the ones corresponding to a displacive transition.

Actually, this conclusion (hidden for many years) is easy to understand when one considers the behavior of the potential energy of the local soft-mode on a simplified way. Basically, the resulting net mechanism is the sum of two contributions: an anharmonic self-energy potential contribution with just a single centered minimum, and a linear contribution coming from the dipole-dipole interaction. Contributions from the other terms are smaller or independent of the local mode so we do not take them into account in this discussion. When the dipole-dipole interaction is added to the self-energy the minimum is displaced from zero toward one of the body diagonals of the unit cell. The diagonal chosen by the system depends on the particular configuration and, due to the symmetry of the problem, all 8-sites may be equally visited. So, instantaneously, the local mode at the unit cell is different from zero as corresponding to an order-disorder mechanism. On the other hand, the averaged potential is the sum of the non-centered single minimum potentials corresponding to all these configurations. Because of the symmetry of the problem at $T>T_c$, contributions coming from the dipole-dipole interaction cancel to each other, and one finally obtains the single local mode self-energy which has a minimum at the center of the unit cell, as corresponding to a displacive transition. Of course, any ferroelectric system having a local mode self-energy with non-centered minima (i.e. with a local cell stable about forming an isolated local mode) should present an strong order-disorder character using any meaningful experimental technique. A further account on this work is given in [15].

Future work on the phase transitions of other $ABO_3$ perovskite like compounds is under consideration.

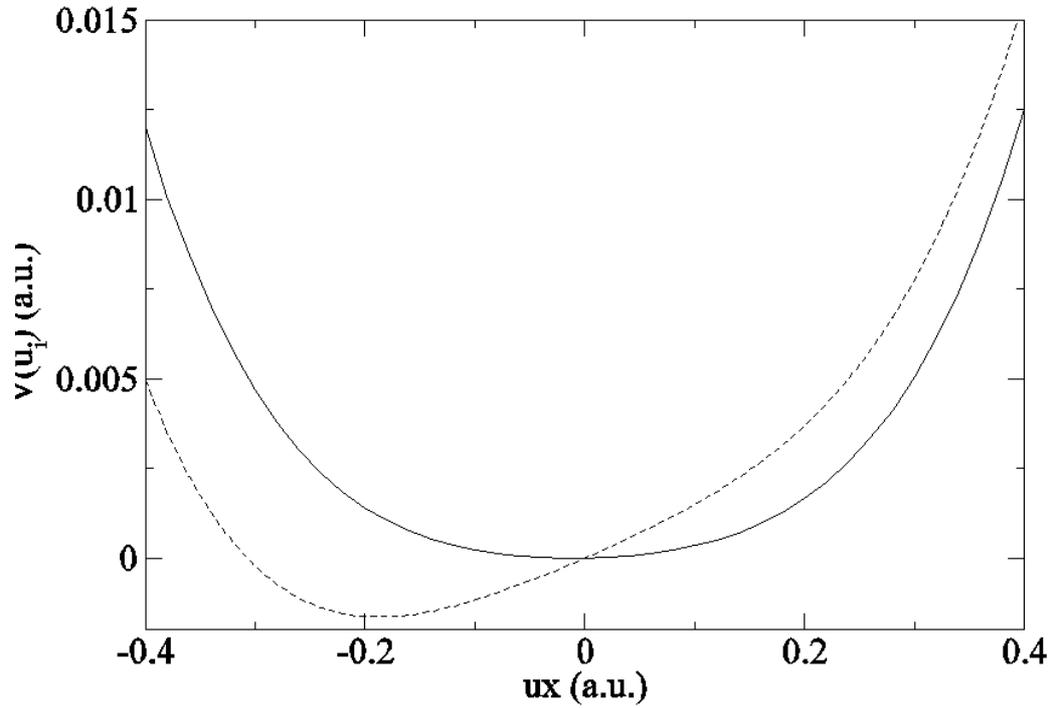

FIGURE 3. Average potential versus $u_i=(ux_i,0,0)$ for T=400K (continuous line) corresponding to the paraelectric phase, and T=250K (dashed line) corresponding to the ferroelectric phase. The average was calculated considering 20000 different configurations.

**ACKNOWLEDGMENTS**

We thank J. Iñiguez, M.G. Stachiotti and R.L. Migoni for useful comments. Financial support from DGCyT is acknowledged.